\documentclass[11pt,a4paper]{article}
\pdfoutput=1
\usepackage{jcappub}
\usepackage[utf8]{inputenc}

\usepackage{color}
\usepackage{amsmath}
\usepackage{pifont}
\usepackage{bbold}

\usepackage{bbm}
\usepackage{verbatim}   
\usepackage{subfigure}
\usepackage{acronym}

\usepackage{amsfonts}
\usepackage{amssymb}
\usepackage{mathtools}
\usepackage{mathrsfs}
\usepackage{graphicx}
\usepackage{multirow}
\usepackage{slashed}
\usepackage{array} 
\newcolumntype{C}{>{\hfil$}p{14pt}<{$\hfil}}
\newcolumntype{Y}{>{\centering\arraybackslash}X}

\usepackage{graphicx}
\usepackage{tabularx}
\usepackage{empheq}
\usepackage{caption}
\usepackage{tikz}
\usetikzlibrary{decorations.pathreplacing}

\newcommand{\diff}[2]{\frac{\mathrm{d}#1 }{\mathrm{d}#2 }}

\title{Primordial Black Hole Constraints with Large Extra Dimensions}

\author[1]{George Johnson}
\emailAdd{george.johnson@physics.ox.ac.uk}

\affiliation[1]{Rudolf Peierls Centre for Theoretical Physics, University of Oxford, OX1 3NP, United Kingdom}

\abstract{We study how the constraints on the primordial black hole density arising from the extragalactic photon background are modified in the scenario that there exist extra large spatial dimensions. We find that though the overall magnitude of the constraints is not substantially different, the mass ranges to which they apply are, and for some choices of mass it is possible for the black holes to constitute the entirety of the dark matter. }

\begin{document}

\maketitle

\section{Introduction}

There is still very little known about the nature of the dark matter in the universe, besides that it is non-relativistic and interacts only weakly with the Standard Model. Whilst it is typically assumed that the dark matter consists of some as-yet undiscovered elementary particle, primordial black holes (PBHs) are a compelling alternative, being naturally cold, dark, and consistent with the framework of known physics. For this reason a great deal of work has been done in understanding the astrophysical and cosmological consequences of a large PBH background; for many choices of PBH mass, strict constraints exist on the fraction of dark matter they could constitute.  See \cite{pbhdm} for a review.

Many of these constraints, in particular those applying to smaller mass PBHs (in the mass range $10^{10}$ g to $10^{17}$ g), are due to the effects of the Hawking radiation these black holes emit. Whilst the existence of Hawking radiation is under little doubt, being necessary for the consistency of a thermodynamic description of black holes, Hawking radiation has never been observed, and so the precise way black holes radiate is still under question. For this reason, it is interesting to consider how modifying the nature of Hawking evaporation modifies the constraints on the density of primordial black holes in the universe.

In this work, we consider how the dominant constraints on the density of small PBHs --- those from the extragalactic gamma ray background (EGB) --- differ in the scenario that black holes can radiate into higher dimensions. The nature of gravity on short scales is not well understood: whilst Coulomb's law (or its quantum field theoretic generalisation) is known to apply down to distances of order $10^{-18}$ m, Newton's law of gravity has only been tested on scales of order several microns. Consequently, it is consistent for there to exist extra large spatial dimensions, and being extraordinarily compact objects, even black holes as heavy as $10^{17}$ g could be sensitive to these dimensions. 

Most new physics, involving the introduction of extra degrees of freedom available for black holes to radiate, would result in little reduction in detectable evaporation products and at best modest weakening of existing constraints. The motivation for studying extra-dimensional evaporation is the critical fact that higher-dimensional black holes, if smaller than the size of the extra dimensions, are significantly \textit{colder} than their 4D counterparts, for a given fixed mass. These PBHs are thus expected to produce fewer evaporation products, and thus be subject to weaker constraints, than ordinary 4D black holes.

This paper is divided into five sections. In Section 2 we review the latest constraints on the PBH density across the entire range of masses. In Section 3 we discuss the behaviour of black holes in theories with large extra dimensions, and explain how the evaporation rate is modified in such a way as to substantially modify the nature of all constraints on low-mass PBHs. In Section 4 we present the main result of our analysis, the modified constraints on the PBH density arising from the extragalactic photon background, for several choices of the number of extra dimensions. Finally, in Section 5, we discuss the modifications we expect to occur to other constraints on the PBH density.

We take $c = \hbar = k_B = 1$ throughout. 

\section{Existing Primordial Black Hole Constraints}

Constraints on the density of primordial black holes can be divided into two classes: constraints from the gravitational effects of the black holes themselves, and constraints from the particles they produce through Hawking evaporation. Since the total rate of energy loss is less for larger black holes, these evaporative constraints exist only for PBHs of mass less than about $10^{17}$ g. Conversely, gravitational effects of PBHs are typically negligible for black holes below this mass.

\begin{figure}
	\centering
	\includegraphics[width=\linewidth]{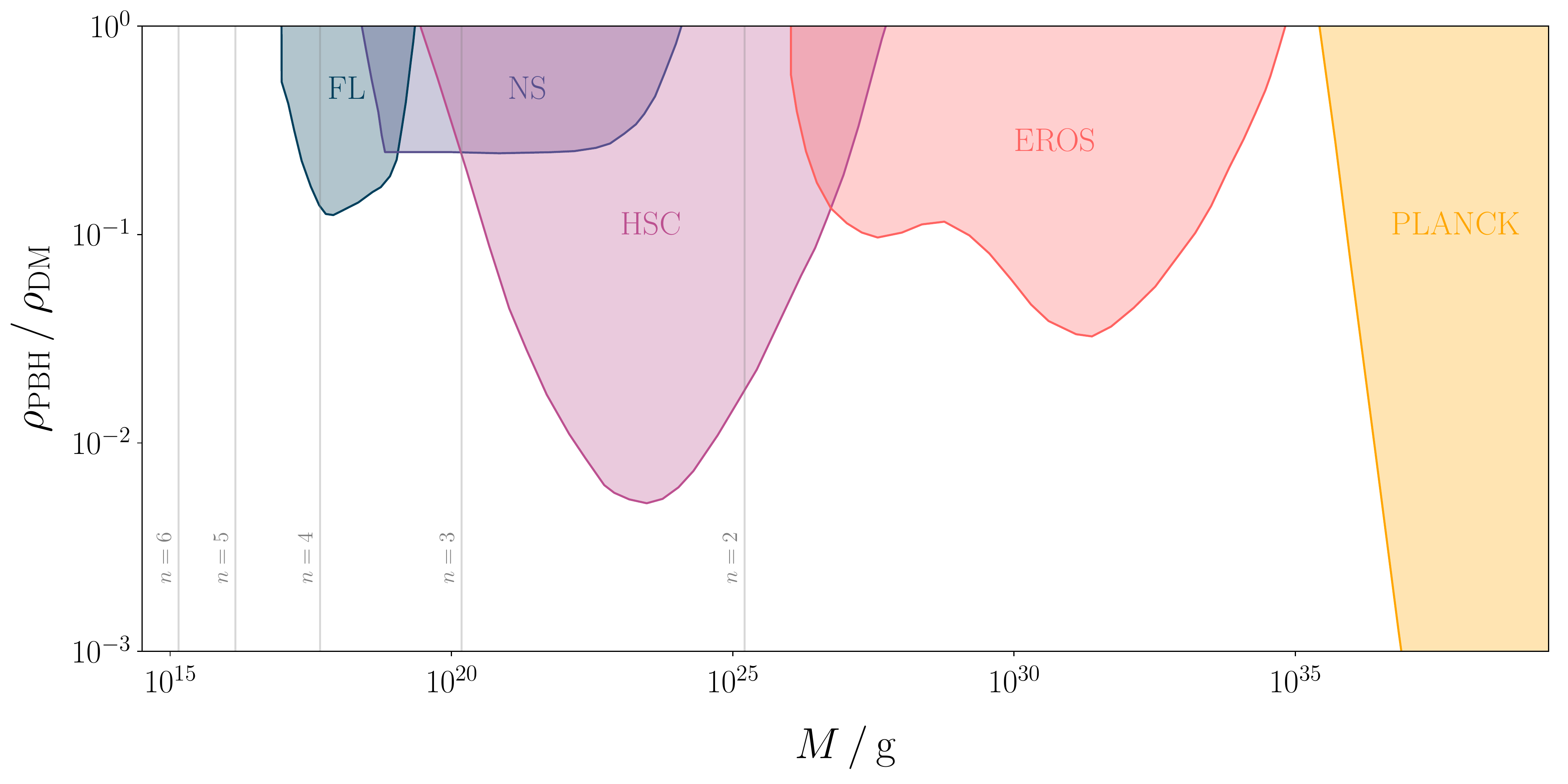}
	
	\includegraphics[width=\linewidth]{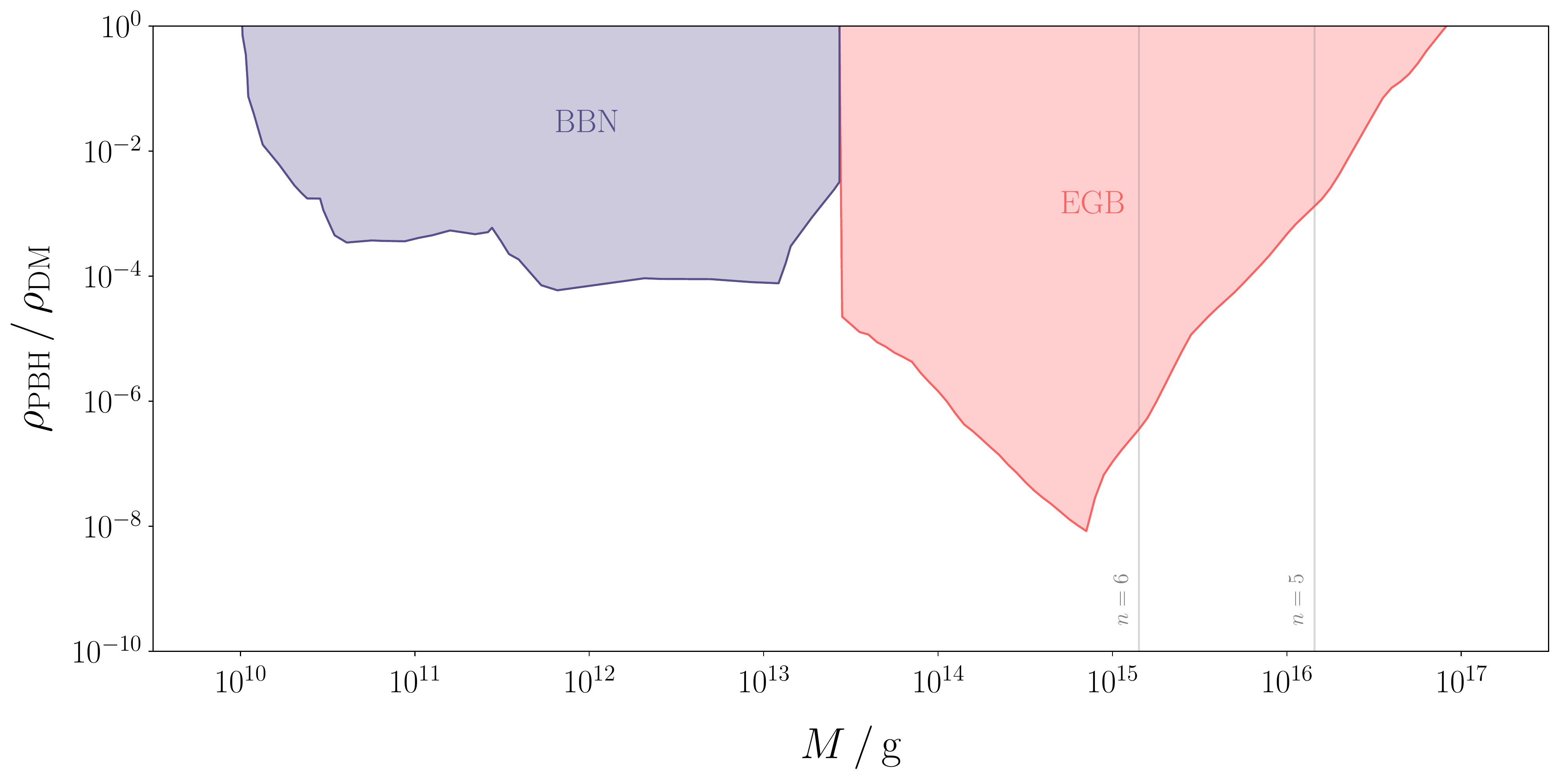}
	\caption{Constraints on the PBH density (at formation) as a fraction of the dark matter density. Shaded regions are excluded. In the upper panel are those constraints arising from the gravitational influence of the black holes, adapted from \cite{extendedmassfunctions}; from left to right the constraints arise from femtolensing experiments (FL), capture by neutron stars (NS), microlensing experiments (HSC and EROS), and effects on the CMB (PLANCK). The vertical grey lines indicate the mass $M_c$ at which the black hole radius is equal to the size of the extra dimensions, as discussed in Section \ref{ledsec}. In the lower panel are those constraints arising from the effects of the PBH evaporation products; the constraints from BBN are taken from \cite{kazconstraints} and those from the EGB we have reproduced in this work.}
	\label{consfig}
\end{figure}

\subsection{Gravitational constraints}

The constraints on the PBH density for masses above around $10^{17}$ g are illustrated in the upper panel of Fig \ref{consfig}. Perhaps the most important of these come from lensing experiments. For small black holes, the wave nature of electromagnetic radiation is significant, and gravitational lensing around the black hole can give rise to an interferometry pattern in the received radiation (termed \textit{femtolensing}) \cite{femto}\footnote{These results have been called into question by \cite{femto2}.}. For larger black holes this interference is not detectable, though gravitational lensing can nevertheless result in the apparent magnification of stars passing behind them (termed \textit{microlensing}) \cite{HSC, EROS}.

There also exist astrophysical effects associated with the collision and subsequent capture of primordial black holes by compact objects such as neutron stars and white dwarfs. In some mass ranges, such collisions would result in the destruction of these objects; the number density of existing neutron stars and white dwarfs hence places mild bounds on the density of primordial black holes \cite{nscapture, whitedwarfs}. 

For large black holes, of order $10^{35}$ g and larger, strict constraints arise from the effects of these PBHs on the CMB. In particular, the accretion of primordial gas around the black holes and subsequent injection of energy into the background plasma would be detectable through its influence on the angular distribution of temperature and polarization of the CMB. Data from Planck strongly constrains this scenario \cite{cmb}.

\subsection{Evaporative constraints}
\label{existingevap}

For masses below about $10^{17}$ g, all constraints on the PBH density are due to the effects of black hole evaporation. The dominant constraints come from the effects on big bang nucleosynthesis (BBN) and the size of the extragalactic gamma ray background (EGB). In \cite{kazconstraints} a comprehensive analysis of these constraints was performed. Injection of high-energy particles during BBN can cause dissociation of heavier isotopes and induce extra interconversion of protons and neutrons, the extent of which is strictly constrained by the known abundances of the light elements. Constraints on black holes that have not fully evaporated today arise from the extragalactic photon background. Continual evaporation of PBHs over the course of the universe's history would have converted a considerable quantity of energy into photons (primarily gamma rays), and this would be observable in the EGB. These constraints are illustrated in the lower panel of Fig. \ref{consfig}.

The constraints from the extragalactic photon background are of most relevance to this work, so we shall endeavour to explain the qualitative form of the constraints. Assuming the black holes are formed very early in the universe, there exists some mass $M_0$ such that they are disappearing today. For black holes much larger than this, incomplete evaporation occurs, and thus not all of the energy in the black holes is converted into Standard Model particles. Since the lifetime of a black hole scales approximately like the cube of its mass, a black hole need not be much heavier than $M_0$ before its lifetime is far longer than the age of the universe and only a small fraction of its energy is converted to photons. The constraints hence weaken as $M$ is increased above $M_0$.

For masses smaller than $M_0$, the black holes have completely disappeared by today. Though the total energy released by the black holes is the same for all such $M$, the smaller the black holes, the earlier they disappeared, and hence the greater the redshifting of the photons they produced. The energy contributed to the photon background today hence decreases as $M$ is decreased below $M_0$, and the constraints weaken. One needs to take a little more care than this --- smaller black holes emit predominantly higher-energy radiation, and the constraints on the size of the photon background are stricter at higher energy. However, sufficient redshifting of frequency occurs that in fact the dominant constraints on small black holes come from the soft end of the gamma ray background. 

For black holes smaller than about $10^{13}$ g, complete evaporation has occurred before photon decoupling. Such radiation is hence not visible in the photon background, and so the constraints disappear completely for such black holes.

We briefly mention that there are several other constraints of an evaporative nature, arising from annihilation of electrons with positrons emitted by the PBHs \cite{positrons, 511}, \textit{galactic} gamma rays and antiprotons  \cite{galacticgamma, galacticantiproton}, and effects on CMB anisotropy \cite{anisotropy}. Apart from over very narrow mass ranges, the constraints from BBN and the EGB are the strictest of these.

\section{Black Holes in Large Extra Dimensions}
\label{ledsec}

The existence of extra compact spatial dimensions is an appealing explanation for the observed weakness of gravity relative to the other fundamental forces \cite{leds}. Informally, the force of gravity is weaker because it is `diluted' amongst these extra dimensions. More formally, in the case that the geometry of the extra dimensions is independent of ours, the gravitational action can be written
\begin{equation} 
S = \frac{1}{16 \pi} M_*^{2+n}\int \, \mathrm{d}^{4+n}x \sqrt{-g} \left( \mathcal{R}_4 + \mathcal{R}_n \right) \,,
\end{equation}
where $M_*$ is the fundamental Planck scale, $\mathcal{R}$ is the Ricci scalar, and $n$ is the number of extra dimensions. Neglecting the second term, we can perform the integral over the extra spatial dimensions to generate an effective 4D action. If $R$ denotes the size of the extra dimensions, we hence find the relation
\begin{equation}
\label{mstar}
M_P^2 \sim M_*^{2+n} R^n \,.
\end{equation}
It is hence possible for the fundamental Planck scale $M_*$ to be far lower than the 4D Planck scale, if the extra dimensions are sufficiently large. For $M_* = 10$ TeV, the above relation implies that $R \sim 10^{11}$ m for $n=1$ --- certainly such a large extra dimension is ruled out by gravitational experiments on the solar system scale. For $n=2$ one finds $R \simeq 25 \, \mu \mathrm{m}$, which is consistent with current short distance tests of Newton's law \cite{torsion}. In this paper we will consider $2 \leq n \leq 6$. In Section \ref{resultssec} we will describe qualitatively the nature of the constraints for $n$ larger than this.

Naturally, experiments exist that test the Standard Model up to around $1 \, \mathrm{TeV}$ --- far smaller distance scales than a micron. If there are additional spatial dimensions, the particles of the Standard Model must not `feel' these extra dimensions --- they must be localised to a brane. The mechanism by which this could occur is an important question, but we note here that it is a natural scenario in the context of some string theories, in which branes occur with gauge theories automatically localised to them.

\subsection{The higher-dimensional Schwarzschild solution }

What is the nature of black holes in this model? Black holes which are much larger in size than these extra dimensions should be insensitive to them, and behave as ordinary 4D black holes. On the other hand, black holes much smaller than these extra dimensions should be insensitive to the finiteness of the dimensions, and behave as $(4+n)$-dimensional objects. There will be some intermediate regime in which the black hole is not adequately described by either picture. However, the crossover between the two descriptions is continuous, for one can show that there is some critical mass $M_c$ at which the size of the extra dimensions, the 4D Schwarzschild radius of the black hole, and the $(4+n)$-dimensional radius of the black hole all approximately coincide. This critical mass is tabulated in Table \ref{tabcrit}. For a review of this material, see \cite{kantireview}.

\begin{table}
	\centering
		\def\arraystretch{1.1}
		\begin{tabularx}{.4\textwidth}{YY}
		\hline
		$n$  & $M_c$ / g  \\
		\hline
		2 & $1.62 \times 10^{25}$ \\
		3 & $1.52 \times 10^{20}$  \\
		4 & $4.65 \times 10^{17}$ \\
		5 & $1.44 \times 10^{16}$ \\
		6 & $1.45 \times 10^{15}$  \\
		\hline
	\end{tabularx}
	\caption{The mass $M_c$ in grams of the black hole whose Schwarzschild radius is equal to the size of the extra dimensions, for $M_* = 10$ TeV.}
	\label{tabcrit}
\end{table}

In $(4+n)$ dimensions the Schwarzschild metric is given by \cite{myersperry}
\begin{equation}
\label{metric}
\mathrm{d}s^2 = -\left(1 - \left(\frac{r_h}{r}\right)^{n+1}\right)\mathrm{d}t^2 + \left(1 - \left(\frac{r_h}{r}\right)^{n+1}\right)^{-1}\mathrm{d}r^2 + r^2 \,\mathrm{d}\Omega_{n+2}^2 \,,
\end{equation}
where the horizon radius is
\begin{equation}
\label{horizonr}
r_h = \frac{1}{M_*}\left(\frac{M}{M_*}\right)^\frac{1}{n+1} \left(\frac{8\, \,\Gamma((n+3)/2)}{(n+2) \pi^{(n+1)/2}}\right)^\frac{1}{n+1} \,.
\end{equation}
The Hawking temperature of such a black hole is given by
\begin{equation}
\label{temperature}
T = M_* \left(\frac{M_*}{M}\right)^\frac{1}{n+1} \left(\frac{n+1}{4 \sqrt{\pi}}\right)\left(\frac{n+2}{8 \Gamma((n+3)/2)}\right)^\frac{1}{n+1} \,.
\end{equation}
The relation between the radius and temperature of the black hole is particularly simple:
\begin{equation}
\label{inversely}
T =  \frac{n+1}{4 \pi r_h} \,.
\end{equation}
When we restrict the metric Eq. \eqref{metric} to the brane, we find a 4D black hole solution whose geometry differs from the ordinary Schwarzschild solution. This will affect the way the black hole gravitates. Consequently, some of the aforementioned gravitational constraints will be modified in this scenario, such as those from lensing experiments, since the bending of light around a black hole is sensitive to the precise geometry. Similarly, capture of these black holes by neutron stars and white dwarfs is sensitive to the potential energy between the two objects, which depends fundamentally on the number of extra dimensions. 

The above notwithstanding, it transpires that for $M_* = 10$ TeV, the critical mass describing the transition from the 4D to the $(4+n)$-dimensional picture occurs close to the mass at which existing constraints become dominantly evaporative\footnote{For $n>2$ at least. For $n=2$, $M_c$ is appreciably larger than this. }. These critical masses are illustrated as vertical lines in Fig. \ref{consfig}. Black holes larger than this critical mass behave as ordinary 4D black holes, and the existing gravitational constraints apply.

On the other hand, Eq. \eqref{temperature} shows that the temperature of a higher-dimensional black hole can differ significantly from that of a 4D Schwarzschild black hole of the same mass. This implies that the rate at which it evaporates, and the energy of the particles it produces during this evaporation, can differ considerably also. We thus expect the constraints on low-mass PBHs to be substantially modified in this scenario.

\subsection{Bulk and brane evaporation}
\label{bulkbrane}

A black hole smaller in size than the extra dimensions radiates gravitons into the bulk and Standard Model particles onto the brane. The radiation of Standard Model particles obeys the usual Stefan-Boltzmann law, with the critical difference that the temperature of the radiation is no longer inversely proportional to the mass of the black hole. As we discuss in the appendix, for both modes of evaporation, the rate of loss of energy is approximately equal, being given by
\begin{equation}
\label{stefanb}
-\diff{M}{t} \sim T^2 \,.
\end{equation}
The differing dependence of the temperature on mass however, gives rise to a black hole lifetime $\tau$ that depends critically on $n$:
\begin{equation}
\label{lifetime}
M_*\tau \sim \left(\frac{M}{M_*}\right)^\frac{n+3}{n+1} \,.
\end{equation}
By considering Eqs. \eqref{horizonr}, \eqref{temperature} and \eqref{lifetime} in conjunction with Eq. \eqref{mstar}, one can show that black holes much smaller than the size of the extra dimensions are larger, cooler, and live longer than 4D black holes of the same mass \cite{millimeter}, at least if one considers emission to involve only a single degree of freedom\footnote{We consider the detail of the number of emitted degrees of freedom in Section \ref{resultssec}.}. However, for black holes of order the size of the extra dimensions, the numerical factors in Eq. \eqref{temperature} are not insignificant, and lead to higher-dimensional black holes disappearing substantially faster.

Throughout the rest of this work we will take $M_* = 10$ TeV. As mentioned at the start of this section, this is approximately the bound for $n=2$ placed on the size of the extra dimensions by measurements of the behaviour of Newton's law on short distances. However, we briefly mention here that there are several other constraints on the size of $M_*$. Firstly, a weak-scale fundamental Planck mass is subject to collider constraints. These are fairly mild, and are consistent with $M_* = 10$ TeV. There are also several astrophysical and cosmological constraints due to the effects of the light Kaluza-Klein modes of the graviton. These bounds are more strict, and typically rule out $M_* = 10$ TeV for $n=2$. However, we note that they are subject to large systematic errors, and depend on the details of the decay of the KK modes. See \S 106 of \cite{pdg} for the latest constraints on $M_*$. In Section \ref{discussion} we describe the qualitative effects of choosing a larger value for $M_*$.

\section{Modified Constraints from the Extragalactic Photon Background}
\label{resultssec}

In this section we present the constraints that exist on the density of higher-dimensional PBHs that arise from the contribution they would make to the extragalactic photon background. It transpires that the strongest constraints come from the X-ray and gamma ray background, as in the 4D case. In the $n=1$ case the black holes radiate primarily in the UV and soft X-ray regions of the electromagnetic spectrum. There are only very poor measurements of the extragalactic UV background, but this is of no consequence since $n=1$ is ruled out by gravitational experiments.

We assume for simplicity a monochromatic mass distribution --- that is, that all of the primordial black holes are formed at the same time with the same mass. This is not particularly realistic, and it is known in the 4D case that constraints tend to become more stringent with extended mass distributions, if qualitatively similar \cite{extendedmassfunctions}. A monochromatic distribution is nevertheless sufficient to indicate the large modifications to the constraints that occur in the extra-dimensional scenario.  We also emphasise that the quantity $\rho_\mathrm{PBH}$ we plot in Fig. \ref{results} is the fraction of the dark matter density the black holes constitute \textit{at formation}, and likewise the mass $M$ is their mass at formation. In order to constitute a fraction of the dark matter today, the PBHs must have an initial mass larger than that mass $M_0$ which would be evaporating now, tabulated in Table \ref{tabevap}.

\subsection{Methodology}

To compute the spectrum of radiation emitted by a black hole, the public code \texttt{BlackHawk} \cite{blackhawk} was used. In its original form, this code computes the emission rate of all Standard Model particles from a given black hole, accounting for greybody factors and using \texttt{PYTHIA} to compute the subsequent decay of all unstable particles. The code makes the simplifying assumption that a black hole begins radiating a given particle only when its temperature exceeds the particle's mass, and thereafter begins radiating it as though it were massless.

The code needs some modification to produce the emission rate in the large extra dimensions scenario. Naturally the mass-radius and mass-temperature relations are modified according to Eqs. \eqref{horizonr} and \eqref{temperature}. Furthermore, the greybody factors differ in different dimensions. These greybody factors were computed for all spins and for all dimensions $n$, using the numerical recipes outlined in \cite{kanti,bulkemission}. Accuracy of the numerical results could be compared to the results in \cite{kanti,gravitonemission2}. We found excellent agreement with the former, although not with the latter. We note that the numerical results in \cite{gravitonemission2} do not agree with the expected low-energy analytic expressions (in particular, all spin-2 greybody factors should go to zero at low energy), so we put the discrepancy between our results down to an error in theirs. To produce the high-energy and low-energy asymptotics of the greybody factors, the analytic results from \cite{JMRbrane1, JMRbrane2} were used.

Given the spectrum of radiation from a black hole at each moment of its lifetime, the density of background photons today (in units of energy per unit volume per unit energy) is given by the formula
\begin{equation}
\label{backgrounddens}
n =  \int_{t_\mathrm{dec}}^{t_\mathrm{max}} (1+z) \frac{\mathrm{d}^2 N}{\mathrm{d}t \, \mathrm{d}E}((1+z)E) \, \mathrm{d}t \,,
\end{equation}
where the derivative in the integrand is the energy being emitted by the black holes per unit time per unit volume per unit energy. The integral is taken between the time of photon decoupling $t_\mathrm{dec}$ and $t_\mathrm{max} = \min(t_0, \tau)$, where $t_0$ is the time today and $\tau$ the lifetime of the black hole. Those photons with energies between $E$ and $E+\mathrm{d}E$, if produced at an earlier time $t$, must have been emitted with blueshifted energy $(1+z)E$ and belonged to a wider energy window $(1+z)\mathrm{d}E$. This accounts for the two redshift factors in the integrand. As appropriate for the matter-dominated era, we take $1+z(t) = (t_0/t)^{2/3}$.

\begin{figure}
	\centering
	\includegraphics[width=0.85\linewidth]{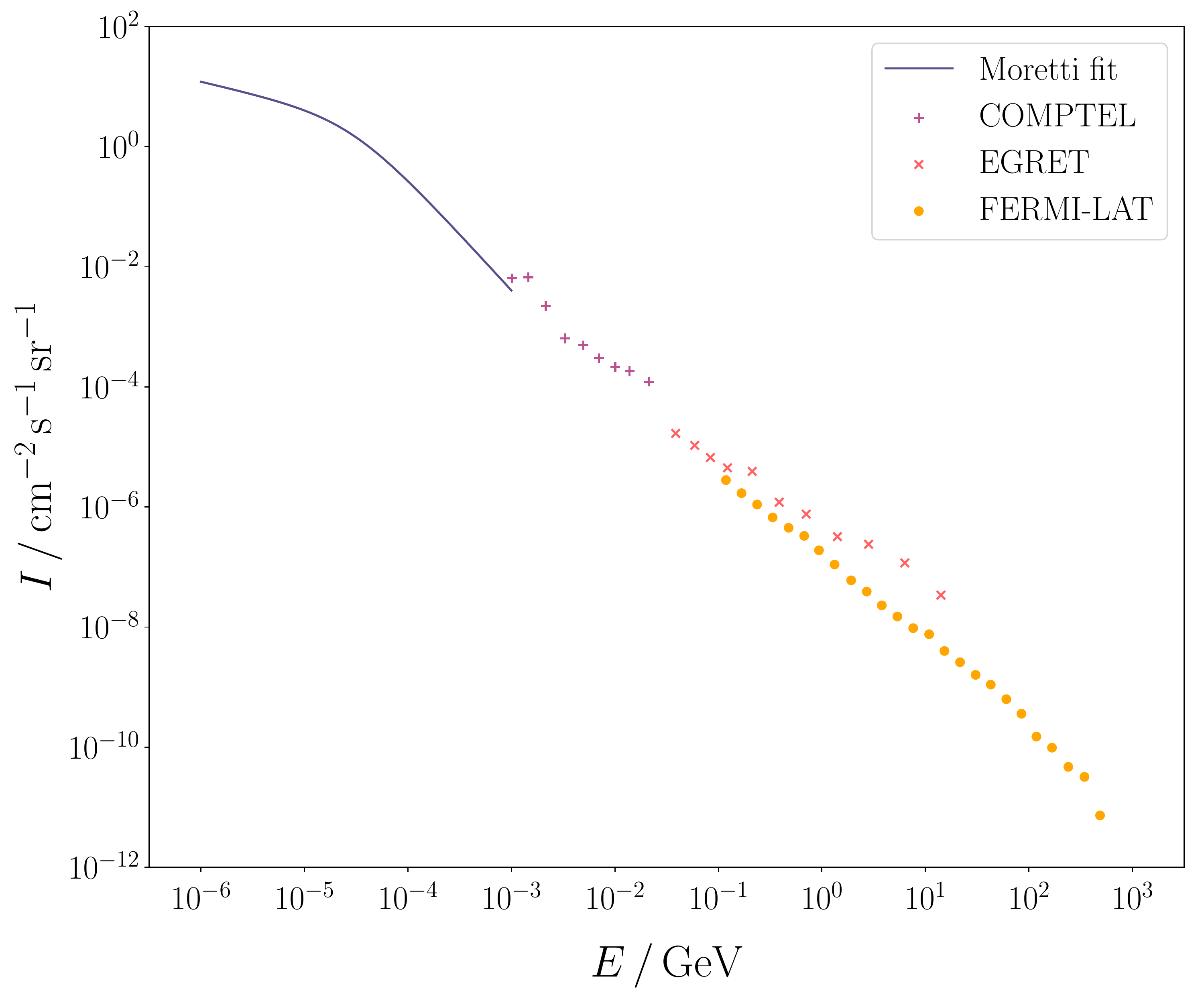}
	\caption{The observed background photon flux in units of energy per square centimetre per second per steradian per unit energy, as a function of energy. The energy range plotted corresponds to the X-ray and gamma ray region of the electromagnetic spectrum. }
	\label{gammadata}
\end{figure}

An isotropic photon background gives rise to an observed flux of
\begin{equation}
I = \frac{c}{4 \pi} n \,.
\end{equation}
The data for the gamma ray region of the electromagnetic spectrum are collected by space-based telescopes, in particular EGRET and COMPTEL aboard the Compton Gamma Ray Observatory and LAT aboard the Fermi Gamma Ray Space Telescope \cite{COMPTEL, EGRET, FERMILAT}.  More data exist for the intensity of the X-ray background, for which we use a fit from \cite{moretti}.  In Fig. \ref{gammadata} we plot these data. In determining the PBH constraints we make the conservative assumption that the entirety of the photon background in the X-ray and gamma ray region of the electromagnetic spectrum is due to black hole evaporation products.

\subsection{Results}
\label{actualresults}

\begin{figure}
	\centering
	\includegraphics[width=\linewidth]{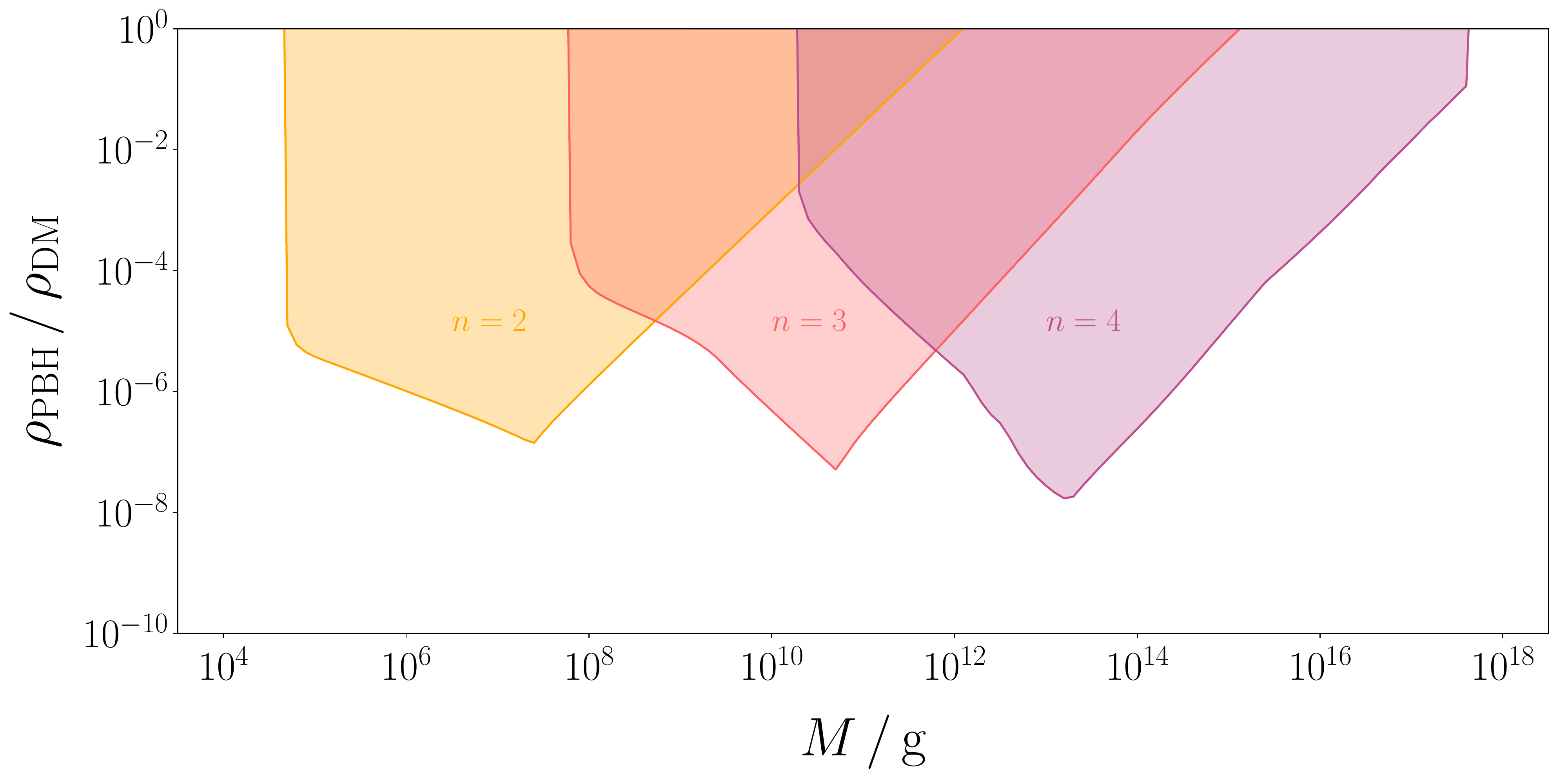}
	
	\includegraphics[width=\linewidth]{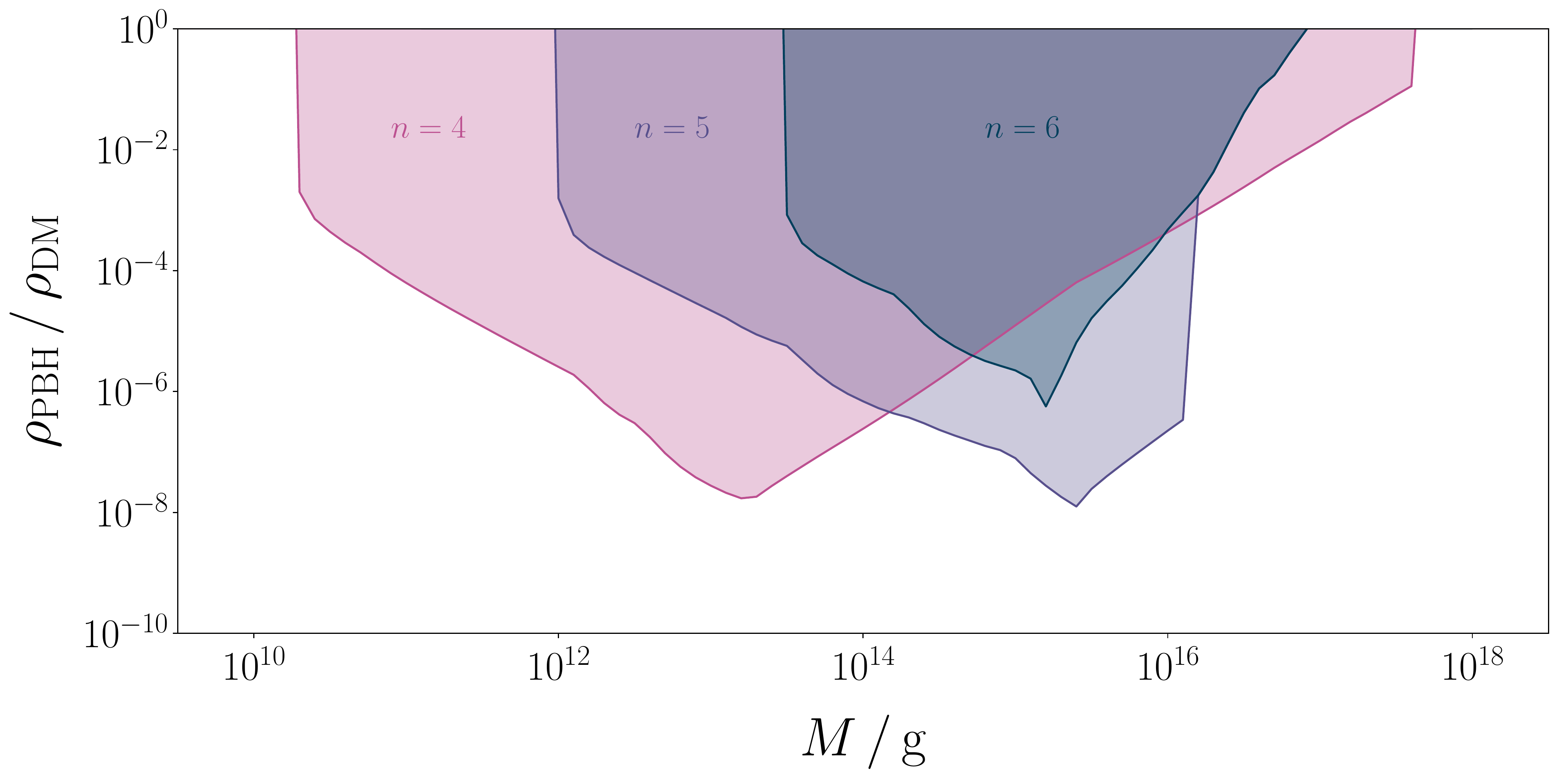}
	
	\caption{Constraints on the PBH density (at formation) as a fraction of the dark matter density for different choices of $n$, assuming a monochromatic mass distribution. Shaded regions are excluded. In the upper panel are plotted (left to right) the constraints for $n=2, 3, 4$ and in the lower panel the constraints for $n=4,5,6$. }
	\label{results}
\end{figure}

Our results are plotted in Fig. \ref{results}. We see that independent of $n$, the shape of the constraints is broadly similar. The explanation for this shape is as outlined in Section \ref{existingevap}: those black holes which are evaporating today contribute most energy to the photon background, with the constraints disappearing for black holes small enough to have evaporated before photon decoupling. The primary qualitative difference is due to the fact that the mass $M_0$ of those black holes evaporating today is dimension-dependent. These masses are tabulated in Table \ref{tabevap}. For $n \geq 4$ we note that the constraints cut off sharply above a certain mass. This is the mass $M_c$ above which the 4D description of the black holes is valid, tabulated in Table \ref{tabcrit}. The total radiation rate is substantially reduced above this mass (as explained in Section \ref{bulkbrane}), which is why the constraints significantly soften. Indeed, comparison with Fig. \ref{consfig} shows that above this mass, the constraints coincide with those in the 4D case. Since we have not treated the crossover behaviour of the black hole solution precisely, the results are not reliable at this mass. 

From Eqs. \eqref{lifetime} and \eqref{temperature} one can show that for a black hole which survives until today, the typical temperature of the Hawking radiation is higher for larger $n$. This explains why the constraints are slightly stronger for larger $n$ (and indeed why they are noisier), the dominant constraints coming from the higher-energy region of the electromagnetic spectrum (see Fig. \ref{gammadata}).  The constraints also cover a wider mass range than the 4D constraints, on account that the lifetime of the black holes Eq. \eqref{lifetime} depends less strongly on mass than in the 4D case.

\begin{table}
	\centering
	\def\arraystretch{1.1}
	\begin{tabularx}{.4\textwidth}{YY}
		\hline
		$n$  & $M_0$ / g  \\
		\hline
		2 & $2.44 \times 10^7$ \\
		3 & $5.33 \times 10^{10}$  \\
		4 & $1.83 \times 10^{13}$ \\
		5 & $2.53 \times 10^{15}$ \\
		6 & $1.45 \times 10^{15}$  \\
		\hline
	\end{tabularx}
	\caption{The mass $M_0$ in grams of a black hole evaporating today, assuming formation at the start of the universe.}
	\label{tabevap}
\end{table}

One might wonder whether radiation into gravitons would dominate photon production for large $n$, and constraints might substantially weaken, on account that the number of degrees of freedom in the graviton is quadratic in $n$ ($g = (n+1)(n+4)/2$). Our results do not bear this hypothesis out. To understand this, we note that the graviton greybody factor is suppressed at low energies relative to those for the photon and neutrinos, and more so for larger $n$. See  Fig. \ref{greybody2} in the appendix.

For $n=7$, the mass $M_c$ is approximately equal to the mass of a 4D black hole evaporating today ($M_0 \simeq 7.09 \times 10^{14}$ g). Black holes smaller than this, for which the extra-dimensional picture is valid, have disappeared by today and therefore could not constitute the dark matter. On the other hand, for black holes larger than this the 4D description, and hence the 4D constraints, apply. Since $M_c$ decreases for larger $n$, we find that for all $n>6$ the constraints on the fraction of the dark matter the black holes could constitute are unchanged from those in the 4D case. 

We note that only for $n=2$ and $n=3$ are there wide mass windows in which PBHs could constitute the entirety of the dark matter --- $M \gtrsim 10^{12}$ g for $n=2$ and $M \gtrsim 10^{15}$ g for $n=3$. For larger $n$ our results show that the photon background places similar constraints on the PBH mass as in the 4D case, requiring $M$ to be larger than about $10^{17}$ g, beyond which other gravitational constraints need to be taken into consideration.

\section{Discussion}
\label{discussion}

The most important question to address next is the modification of the constraints from BBN in the extra-dimensional scenario. Just as for the photon background, we expect that the dominant constraints will arise from those black holes which evaporate completely during BBN. Since such black holes will be lighter than their 4D counterparts, we expect the constraints from BBN to be shifted to lower mass, and more so for lower $n$. We leave a detailed study of this to future work.

It is also necessary to understand how other evaporative and gravitational constraints change with the introduction of extra dimensions. As mentioned in Section \ref{actualresults}, the typical temperature of the black holes increases as $n$ is increased. For low $n$, the only evaporation products are neutrinos, gravitons, and photons, and we hence expect the constraints from positron annihilation or antiprotons to be non-existent. For large $n$, black holes larger than $10^{17} g$ behave as 4D objects, and so we expect existing gravitational constraints to apply. Lensing constraints in the low $n$ case, and other evaporative constraints in the large $n$ case, would need to be studied in greater detail.

The purpose of this work has been to demonstrate the significant qualitative changes to the constraints on the PBH density that occur when extra dimensions are present. In doing so we have made some simplifying assumptions --- that the black hole mass distribution is monochromatic, and that the black holes are not rotating. In any number of dimensions, the Hawking temperature of a black hole depends quite sensitively on its angular momentum; the constraints on a population of spinning black holes could thus be appreciably different. To understand how these two factors affect the constraints in the 4D case, see \cite{rotatingconstraints}.

A final natural question to ask is how the constraints would differ for a different choice of $M_*$. In this work we have chosen the lowest value of $M_*$ consistent with experiment. From Eq. \eqref{lifetime} we see that a black hole evaporating today would have larger mass for a larger choice of $M_*$. We thus expect the constraints to be shifted to larger mass as $M_*$ is increased, though with the same qualitative shape. For $M_* = 10$ TeV we find that the constraints for $n > 6$ are just as in the 4D case, and we expect that the larger we take $M_*$, the fewer choices of $n$ will give rise to novel constraints. 

\section*{Acknowledgements}
I am grateful to John March-Russell for many useful discussions, and for initially suggesting the possibility that constraints on the PBH density could differ depending on the number of dimensions the black holes exist in.

\appendix

\section{Radiation in Higher Dimensions}

In this appendix we state the formulae for computing the total emission rate
\begin{equation}
\frac{\mathrm{d}^2 N}{\mathrm{d}t \, \mathrm{d}\omega} \,,
\end{equation}
as appears in Eq. \eqref{backgrounddens}, for both bulk and brane emission. The most straightforward expression, which applies in any number of dimensions, is
\begin{equation}
\label{simple}
\frac{\mathrm{d}^2 N}{\mathrm{d}t \, \mathrm{d} \omega} =  \frac{1}{2 \pi} \sum_\text{states} \frac{|\mathcal{A}(\omega)|^2}{\exp(\omega/T) \mp 1} \,,
\end{equation}
where $|\mathcal{A}|^2$ is the absorption probability for a particle incident upon the black hole. A formula that makes closer connection with the usual Planck law is
\begin{equation}
\label{closer}
\frac{\mathrm{d}^2 N}{\mathrm{d}t \, \mathrm{d} \omega} = \frac{g\,\Omega_{n+2}}{(2\pi)^{n+3}} \omega^{n+2}   \frac{\sigma(\omega)}{\exp(\omega/T)\mp 1} \,,
\end{equation}
where $g$ is the number of degrees of freedom of the particle. Here the quantity $(4+n)$ describes the number of dimensions the emitted particles feel --- even for a higher-dimensional black hole, we should take $n=0$ for brane emission. These two expressions can be seen to be equivalent using the relation between the greybody factor and the absorption amplitude:
\begin{equation}
\label{gbemission}
\sigma(\omega) = \frac{2^n \pi^{(n+1)/2}(n+1)\Gamma((n+1)/2)}{\omega^{n+2}} \sum_\ell N_\ell |\mathcal{A}_\ell (\omega)|^2 \,,
\end{equation}
where $N_\ell$ is the number of degrees of freedom per angular momentum mode $\ell$, implicit in the sum over states in Eq. \eqref{simple}. In four dimensions we have $N_\ell = 2\ell+1$, though this is no longer true in higher dimensions, and indeed $N_\ell$ depends on whether the degree of freedom being emitted is a scalar, vector, or tensor perturbation.  We remark that Eq. (2.8) in \cite{kanti} is not correct in general since it assumes the scalar form of $N_\ell$.

In general the greybody factor has only weak dependence on frequency. In particular, for large energies the greybody factor asymptotes to a constant, the absorption cross-section of the black hole. Note that for a perfect black body, the greybody factor is independent of $\omega$ and equal to the cross-sectional `area' of the body, $\pi r^2$ for a 4D black hole. In this case we can multiply Eq. \eqref{closer} by energy and integrate to produce the total rate of radiation. It is particularly simple:
\begin{equation}
\label{etazeta}
P = \frac{g(n+3)}{\pi}\, T^{n+4}\, r_h^{n+2}	\begin{cases}
	\zeta(n+4) & \mathrm{bosons} \,, \\
	\eta(n+4) & \mathrm{fermions} \,,
	\end{cases}
\end{equation}
where $\zeta(z)$ is the Riemann zeta function and $\eta(z)$ the Dirichlet eta function, both approximately unity for large $n$.  To an order of magnitude, we thus have
\begin{equation}
P \sim  T^{n+4} \, r_h^{n+2} \,.
\end{equation}
Since the horizon radius and Hawking temperature are inversely proportional, as in Eq. \eqref{inversely}, we arrive at the conclusion that the rate of loss of mass of the black hole is always proportional to $T^2$, independent of the number of dimensions, as claimed in Eq. \eqref{stefanb}. Studying Eq. \eqref{etazeta} in more detail, we can conclude that for a given $n$ the rates of bulk and brane emission are of the same order of magnitude. However, note that the numerical factors in the expression for the temperature Eq. \eqref{temperature} are not unimportant, and result in orders-of-magnitude faster evaporation for $n=6$ as compared to $n=1$. For this reason, constraints on the PBH density substantially weaken once the radius of the black hole exceeds the size of the extra dimensions.

For black holes larger than the size of the extra dimensions, the spectrum contains the 4D graviton and a tower of Kaluza-Klein modes, which the black hole can also radiate in principle. However, the lightest of these has mass of order $1/R$, which by Eq. \eqref{inversely} only begins to be radiated when $r_h \sim R$. Such radiation can hence be neglected for large black holes, and evaporation treated just as in the 4D case. The exact spectrum of Kaluza-Klein modes depends on the geometry of the extra dimensions, but Weyl's law guarantees that as the black hole becomes much smaller than the size of the extra dimensions, the spectrum tends towards that for infinite space. 

We briefly mention that conventional Kaluza-Klein dimensional reduction does give rise to other massless modes, arising from the symmetries of the extra dimensions, such as the moduli fields which describe the dimensions’ size. For both phenomenological and theoretical reasons such fields must by some mechanism gain a mass: this evades the significant experimental
implications of additional massless particles, and stabilises the size and shape of the extra dimensions. We have assumed in this work that this moduli problem is solved in such a way as to raise the masses of these modes at least as high as that of the lightest massive Kaluza-Klein mode. This ensures that for black holes larger than the extra dimensions, ordinary 4D evaporation applies.

\clearpage

\begin{figure}[h]
	\includegraphics[width=0.95\linewidth]{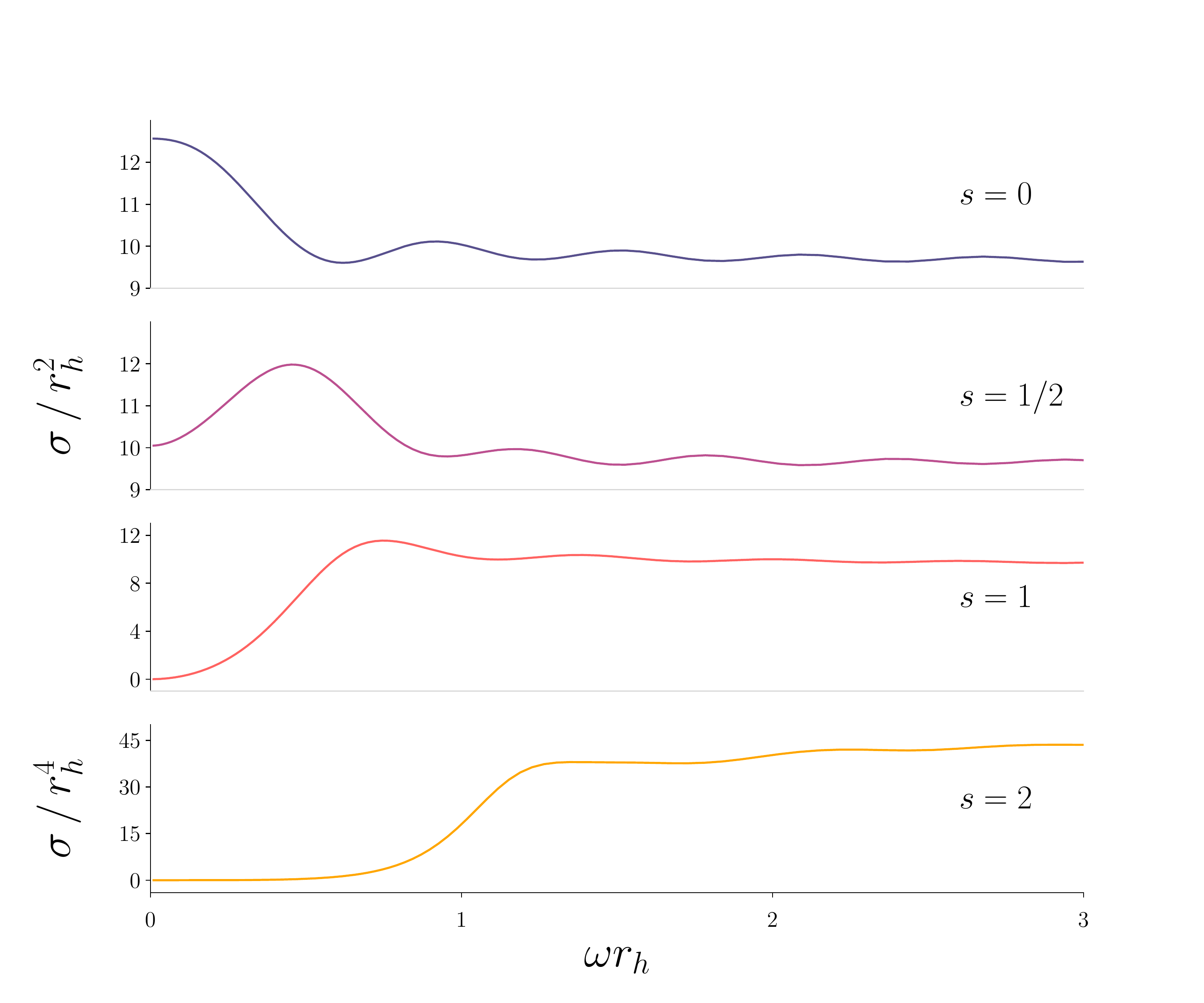}
	\caption{Sample greybody factors for $n=2$, for all types of particle. Note the different scales on the $y$-axes, and the low-energy suppression of graviton radiation. }
	\label{greybody2}
\end{figure}

\bibliography{dims}
\bibliographystyle{JHEP}

\end{document}